\documentclass[english,aps,prl,twocolumn]{revtex4}
\usepackage[T1]{fontenc}
\usepackage[latin1]{inputenc}
\usepackage{graphicx}

\makeatletter

\providecommand{\LyX}{L\kern-.1667em\lower.25em\hbox{Y}\kern-.125emX\@}

\usepackage{babel}
\makeatother
\begin{document}

\preprint{This line only printed with preprint option}

\title{Acoustical-Mode-Driven Electron-Phonon Coupling in Transition-Metal
Diborides}

\author{Prabhakar P. Singh}

\email{ppsingh@phy.iitb.ac.in}

\affiliation{Department of Physics, Indian Institute of Technology, Powai, Mumbai-
400076, India}

\begin{abstract}
We show that the electron-phonon coupling, $\lambda _{\mathbf{q}\nu }$,
in the transition-metal diborides $NbB_{2}$ and $TaB_{2}$ is dominated
by the longitudinal acoustical ($LA)$ mode, in contrast to the optical
$E_{2g}$ mode dominated coupling in $MgB_{2}.$ Our \emph{ab initio}
results, described in terms of phonon dispersion, linewidth, and $\lambda _{\mathbf{q}\nu }$
along $\Gamma -A$, also show that (i) $NbB_{2}$ and $TaB_{2}$ have
a relatively weak electron-phonon coupling, (ii) the $E_{2g}$ linewidth
is an order of magnitude larger in $MgB_{2}$ than in $NbB_{2}$ or
$TaB_{2}$, (iii) the $E_{2g}$ frequency in $NbB_{2}$ and $TaB_{2}$
is considerably higher than in $MgB_{2}$, and (iv) the $LA$ frequency
at $A$ for $TaB_{2}$ is almost half of that of $MgB_{2}$ or $NbB_{2}$.
\end{abstract}

\pacs{63.20.Kr,74.25.Kc}

\maketitle
The discovery of superconductivity in $MgB_{2}$ \cite{akimitsu,nagamatsu}
has renewed interests \cite{young,yamamoto,akimitsu2,Kaczorowski,muzzy,gasparov,souptel,bharathi,note1}
in finding superconductivity in similar materials such as simple metal
diborides ($BeB_{2},\, BeB_{2.75}$ \cite{young}), transition-metal
diborides ($NbB_{2}$ \cite{yamamoto,akimitsu2}, $TaB_{2}$ \cite{yamamoto,Kaczorowski},
$MoB_{2}$ \cite{muzzy}, $ZrB_{2}$ \cite{gasparov}), borocarbides
($LiBC$ \cite{souptel,bharathi}), and other alloys \cite{note1}.
Surprisingly, many of the materials mentioned above have not shown
any superconductivity, those who are found to superconduct do so at
a relatively low $T_{c}<10\, K,$ and very likely, require hole doping
and/or external pressure to show superconductivity. For example, the
recent work of Yamamoto \emph{et al}. \cite{yamamoto} \emph{}showing
superconductivity in hole-doped $Nb_{x}B_{2}$ is a case in point.
Thus, of all the diborides, $MgB_{2}$ seems to be in a class by itself
with a $T_{c}=39\, K$while all the other diborides have $T_{c}<10\, K.$

Recent work on $MgB_{2}$ \cite{kortus,an,pps_prl,kong,bohnen,yildirim,choi1,shukla,pps_nbb2}
have unambiguously shown a very strong and anisotropic electron-phonon
coupling in this system. In particular, it is found that electron/hole
states on the cylindrical sheets of the Fermi surface along $\Gamma -A$
couple very strongly to the in-plane $B-B$ bond stretching $E_{2g}$
phonon mode \cite{kong,bohnen,yildirim,choi1,shukla}. The $E_{2g}$
phonon mode coupling gives rise to partial electron-phonon coupling
constant $\lambda _{\mathbf{q}\nu }$ of the order of $2-3$ \cite{shukla}.
Such a large partial $\lambda _{\mathbf{q}\nu }$ along $\Gamma -A$
results in a relatively high superconducting transition temperature
of $39\, K.$ Thus, the observed differences in the superconducting
properties of $MgB_{2}$ \emph{vis-a-vis} other diborides, specially
the transition-metal diborides, can be better understood by comparing
the electron-phonon coupling along $\Gamma -A$ in these systems.
Taking the cue, we have studied from first principles (i) the phonon
dispersion $\omega _{\mathbf{q}\nu }$, (ii) the phonon linewidth
$\gamma _{\mathbf{q}\nu }$ \cite{allen1}, and (iii) the partial
electron-phonon coupling constant $\lambda _{\mathbf{q}\nu }$ along
$\Gamma -A$ in $MgB_{2},\, NbB_{2}$ and $TaB_{2}$ in $P6/mmm$
crystal structure. 

We have calculated the electronic structure of $MgB_{2},\, NbB_{2}$
and $TaB_{2}$ in $P6/mmm$ crystal structure. We used optimized lattice
constants $a$ and $c$ for $MgB_{2}$ and $NbB_{2}$ \cite{pps_nbb2},
and experimental lattice constants ($a=5.826\, a.u.,$ $c=6.130\, a.u.$)
for $TaB_{2}$ which are close to our optimized values of $a=5.792\, a.u.$
and $c=6.143\, a.u.$. The lattice constants were optimized using
the ABINIT program \cite{abinit} based on pseudopotentials and plane
waves. For studying the electron-phonon interaction we used the full-potential
linear response program of Savrasov \cite{savrasov1,savrasov2} to
calculate the dynamical matrices and the Hopfield parameter, which
were then used to calculate the phonon dispersion $\omega _{\mathbf{q}\nu }$,
the phonon linewidth $\gamma _{\mathbf{q}\nu }$, and the partial
electron-phonon coupling constant $\lambda _{\mathbf{q}\nu }$ along
$\Gamma -A$ in $MgB_{2},\, NbB_{2}$ and $TaB_{2}$.

Based on our calculations, described below, we find that (i) in contrast
to a strong and $E_{2g}$-mode dominated electron-phonon coupling
in $MgB_{2}$, the transition-metal diborides $NbB_{2}$ and $TaB_{2}$
have a relatively weak electron-phonon coupling which is dominated
by the longitudinal acoustical $(LA)$ mode, (ii) the $E_{2g}$ phonon
linewidth is an order of magnitude larger in $MgB_{2}$ than in $NbB_{2}$
or $TaB_{2}$, and (iii) the $E_{2g}$ phonon frequency in $NbB_{2}$
as well as $TaB_{2}$ is considerably higher than in $MgB_{2}$ while
the $LA$ phonon frequency at $A$ for $TaB_{2}$ is almost half of
that of $MgB_{2}$ or $NbB_{2}$. 

Before describing our results in detail, we provide some of the computational
details. As indicated above, the structural relaxation was carried
out by the molecular dynamics program ABINIT \cite{abinit} with Broyden-Fletcher-Goldfarb-Shanno
minimization technique using Troullier-Martins pseudopotential \cite{troullier}
for $MgB_{2}$ and Hartwigsen-Goedecker-Hutter pseudopotentials \cite{hgh}
for $NbB_{2}$ and $TaB_{2}$, 512 Monkhorst-Pack \cite{monkhorst}
$\mathbf{k}$-points and Teter parameterization for exchange-correlation.
The kinetic energy cutoff for the plane waves was $110\, Ry$ for
$MgB_{2}$ and $140\, Ry$ for $NbB_{2}$ and $TaB_{2}.$ The charge
self-consistent full-potential LMTO \cite{savrasov1} calculations
for electronic structure were carried out with the generalized gradient
approximation for exchange-correlation of Perdew \emph{et al} \cite{perdew}
and 484 $\mathbf{k}$-points in the irreducible wedge of the Brillouin
zone. In these calculations, we used $s,$ $p,$ $d$ and $f$ orbitals
at the $Mg$ and $Ta$ sites, and $s,$ $p$ and $d$ orbitals at
the $Nb$ and $B$ sites. The $2p$ state of $Mg$ as well as the
$5s$ and $5p$ states of $Ta$ were treated as semi-core states.
In all cases the potential and the wave function were expanded up
to $l_{max}=6$. The muffin-tin radii for $Mg$, $B$, $Nb$ and $Ta$
were taken to be $2.4,$ $1.66,$ $2.3$ and $2.5$ atomic units,
respectively. 

\begin{figure}
\begin{center}\includegraphics[  width=7.4cm,
  height=7.4cm,
  angle=270,
  origin=lB]{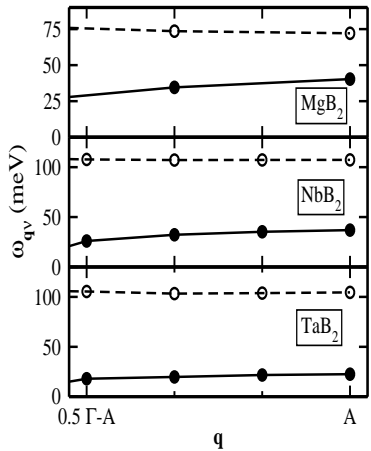}\end{center}

\caption{The phonon dispersion $\omega _{\mathbf{q}\nu }$ of $MgB_{2}$ (upper
panel), $NbB_{2}$ (middle panel) and $TaB_{2}$ (lower panel) for
longitudinal acoustical (solid circles connected with full line) and
optical $E_{2g}$ (open circles connected with dashed line) modes
along $\Gamma -A$, calculated using the full-potential linear response
method as described in the text. The lines connecting the points are
only a guide to the eye. }
\end{figure}

The calculation of dynamical matrices and the Hopfield parameters
along $\Gamma -A$ were carried out for $4$ equidistant $\mathbf{q}$-points
for $MgB_{2}$ and $7$ equidistant $\mathbf{q}$-points for $NbB_{2}$
and $TaB_{2}.$ For Brillouin zone integrations we used a $12\times 12\times 12$
grid while the Fermi surface was sampled more accurately with a $36\times 36\times 36$
grid of $\mathbf{k}$-points using the double grid technique as outlined
in Ref. \cite{savrasov2}. We checked the convergence of the relevant
quantities by carrying out Brillouin zone integrations using a $16\times 16\times 16$
grid of $\mathbf{k}$-points with Fermi surface sampling done over
$48\times 48\times 48$ grid.

\begin{table}

\caption{The phonon frequencies (in $meV$) at $\Gamma $ and $A$ for the
$E_{2g}$ and the $LA$ modes for $MgB_{2}$ calculated using the
linear response method as described in the text as well as from previous
work. }

\begin{tabular}{|c|c|c|c|}
\hline 
&
$\omega _{\Gamma (E_{2g})}$&
$\omega _{A(E_{2g})}$&
$\omega _{A(LA)}$\\
\hline
\hline 
Present work&
78&
72&
40\\
\hline 
Kong \emph{et al.} \cite{kong}&
73&
71&
38\\
\hline 
Bohnen \emph{et al.} \cite{bohnen}&
71&
63&
39\\
\hline 
Shukla \emph{et al.} \cite{shukla}&
65&
57&
37\\
\hline
\end{tabular}
\end{table}

In Fig. 1 we show the phonon dispersion of $MgB_{2}$, $NbB_{2}$
and $TaB_{2}$ for $LA$ and optical $E_{2g}$ modes along $\Gamma -A$.
For $MgB_{2}$, a comparison of our calculations of $E_{2g}$ and
$LA$ phonon frequencies at $\Gamma $ and $A$ points with the previous
calculations of Refs. \cite{kong,bohnen,shukla} is given in Table
I. As expected, our results are closer to the work of Ref. \cite{kong}.
The difference in the $E_{2g}$ phonon frequency at $\Gamma $, which
has been found to be very sensitive to the structural input and Brillouin
zone integration \cite{bohnen}, arises due to the experimental lattice
constants used in Refs. \cite{kong,bohnen,shukla}. The value calculated
by Shukla \emph{et al.} for $E_{2g}$ phonon frequency is somewhat
lower. However, our calculated frequency for $LA$ phonon mode at
$A$ is in good agreement with previous calculations. We expect $\omega _{\mathbf{q}\nu }$
for $NbB_{2}$ and $TaB_{2}$ to have similar accuracy. 

In $NbB_{2}$ and $TaB_{2}$, the $E_{2g}$ phonon mode along $\Gamma -A$
has considerably stiffened in comparison with $MgB_{2}$. The $E_{2g}$
frequency in $MgB_{2}$ changes from $78$ to $72\, meV$, while for
$NbB_{2}$ and $TaB_{2}$ it changes from $110$ to $107\, meV$ and
$108$ to $105\, meV$ from $\Gamma $ to $A$, respectively. The
$E_{2g}$ frequency at $\Gamma $ in $TaB_{2}$, as calculated by
Rosner \emph{et al.} \cite{rosner}, is $98\, meV.$ The $LA$ phonon
mode at $A$ for $TaB_{2}$ $(22\, meV)$ is almost half of that of
$MgB_{2}$ $(40\, meV)$ and $NbB_{2}$ $(37\, meV)$. Here, we like
to point out that in the present work the number of \textbf{$\mathbf{q}$}
points chosen along $\Gamma -A$ is not sufficient to say anything
about the anomaly in the acoustical mode.

\begin{figure}
\begin{center}\includegraphics[  width=7.4cm,
  height=7.4cm,
  angle=270,
  origin=lB]{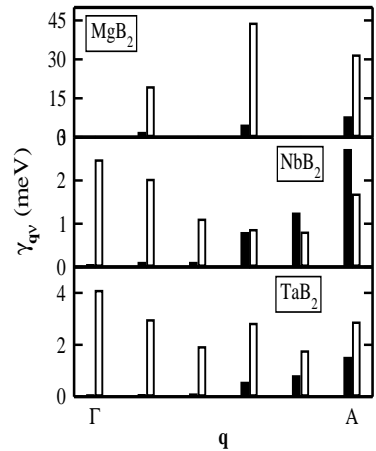}\end{center}

\caption{The phonon linewidth $\gamma _{\mathbf{q}\nu }$ of $MgB_{2}$ (upper
panel), $NbB_{2}$ (middle panel) and $TaB_{2}$ (lower panel) for
$LA$ (solid bar) and optical $E_{2g}$ (open bar) modes along $\Gamma -A$,
calculated using the full-potential linear response method as described
in the text. The phonon linewidth for the $LA$ mode in $MgB_{2}$
has been multiplied by a factor of 10 for clarity.}
\end{figure}

The differences in the nature of electron-phonon interaction between
$MgB_{2}$ and the transition-metal diborides $NbB_{2}$ and $TaB_{2}$
become quite apparent if one considers the electron-phonon contribution
to the phonon lifetimes. In the case of $MgB_{2}$, as shown by Shukla
\emph{et al.} \cite{shukla}, the anharmonic effects \cite{yildirim,choi1}
make negligible contribution to the the phonon linewidth . Thus, the
anomalous broadening of the $E_{2g}$ phonon linewidth along $\Gamma -A$
underscores the strength of the electron-phonon coupling for this
particular mode in $MgB_{2}$ \cite{shukla}. To see what happens
in the transition-metal diborides, we show in Fig. 2 the phonon linewidths
of $MgB_{2}$, $NbB_{2}$ and $TaB_{2}$ for $LA$ and $E_{2g}$ modes
along $\Gamma -A$. For $MgB_{2}$, our calculated $\gamma _{\mathbf{q}\nu }$'s
are in reasonable agreement with the results of Ref. \cite{shukla}.
Note that the values shown in Fig. 2 correspond to twice the linewidth.
From Fig. 2, it is clear that in $MgB_{2}$ the electron-phonon coupling
along $\Gamma -A$ is dominated by the optical $E_{2g}$ phonon mode
with a maximum $\gamma _{E_{2g}}$ of $44\, meV$, and that the $LA$
mode plays essentially no role. In contrast, in $NbB_{2}$ and $TaB_{2}$
(i) the linewidths are more than an order of magnitude smaller than
in $MgB_{2}$, \emph{}for example the maximum $\gamma _{E_{2g}}$
is only about $4\, meV$ in $TaB_{2}$ and (ii) the contribution from
the $E_{2g}$ mode decreases from $4$ to $2.8\, meV$, while that
due to $LA$ mode increases from $0.02$ to $1.5\, meV$, as one moves
from $\Gamma $ to $A$. The phonon linewidths of $MgB_{2}$ and the
transition-metal diborides $NbB_{2}$ and $TaB_{2}$, as described
above, clearly demonstrate the differences in the strength and the
nature of electron-phonon interaction in these systems. 

\begin{figure}
\begin{center}\includegraphics[  width=7.4cm,
  height=7.4cm,
  angle=270,
  origin=lB]{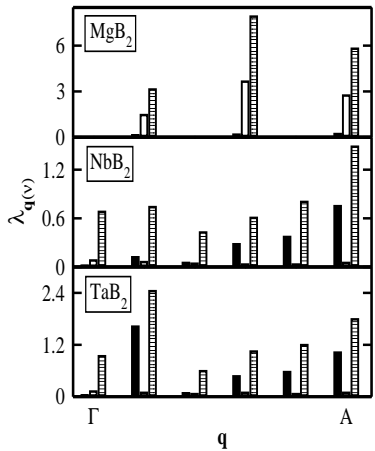}\end{center}

\caption{The partial electron-phonon coupling constant $\lambda _{\mathbf{q}\nu }$
of $MgB_{2}$ (upper panel), $NbB_{2}$ (middle panel) and $TaB_{2}$
(lower panel) for $LA$ (solid bar) and optical $E_{2g}$ (open bar)
modes along $\Gamma -A$, calculated using the full-potential linear
response method as described in the text. The total electron-phonon
coupling constant $\lambda _{\mathbf{q}}$ (lined bar) is also shown.}
\end{figure}

To see the strengths with which the $LA$ and the $E_{2g}$ phonon
modes couple to the electrons, we show in Fig. 3 the partial as well
as the total electron-phonon coupling constant $(\lambda _{\mathbf{q}})$
along $\Gamma -A$ for $MgB_{2},\, NbB_{2}$ and $TaB_{2}$. Certainly,
the most striking feature of these systems, as evidenced in Fig. 3,
is the overall strength of the electron-phonon coupling in $MgB_{2}$
$(\lambda _{\mathbf{q}}\sim 7.9)$ as compared to $NbB_{2}$ $(\lambda _{\mathbf{q}}\sim 1.5)$
and $TaB_{2}$ $(\lambda _{\mathbf{q}}\sim 2.5)$; nevertheless the
additional feature of $E_{2g}-$dominated $\lambda _{\mathbf{q}}$
in $MgB_{2}$ giving way to $LA$-dominated $\lambda _{\mathbf{q}}$
in $NbB_{2}$ and $TaB_{2}$ is just as striking. Thus, the electron-phonon
coupling in the transition-metal diborides $NbB_{2}$ and $TaB_{2}$
is essentially due to the $LA$ mode (the transverse acoustical mode
makes some contribution), the contribution from the $E_{2g}$ mode
being insignificant. For $TaB_{2}$, Rosner \emph{et al.} \cite{rosner}
also found $\lambda _{E_{2g}}=0.05$ at $\Gamma $, in agreement with
the present work. However, their \cite{rosner} conclusion about the
strength of the electron-phonon coupling in $TaB_{2}$ is erroneous
because it doesn't take into account the contributions from the acoustical
modes properly. We also note that, in our opinion \cite{pps_nbb2,pps_unp},
the experimentally \cite{naiduk} deduced electron-phonon coupling
in $NbB_{2}$ and $TaB_{2}$ are underestimated. These differences
in the electron-phonon coupling between $MgB_{2}$ and the transition-metal
diborides $NbB_{2}$ and $TaB_{2}$ may help explain why $MgB_{2}$
superconducts at $39\, K$ while $NbB_{2}$ and $TaB_{2}$ do not
show any superconductivity down to $2\, K$. 

\begin{figure}
\begin{center}\includegraphics[  width=7.4cm,
  height=7.4cm,
  angle=270,
  origin=lB]{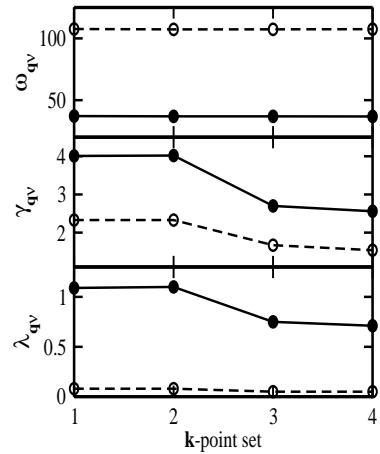}\end{center}

\caption{The convergence of $\omega _{\mathbf{q}\nu }$, $\gamma _{\mathbf{q}\nu }$,
and $\lambda _{\mathbf{q}\nu }$ at $A$ in the Brillouin zone for
$NbB_{2}$ as a function of $\mathbf{k}$-points using the double-grid
technique as described in the text. In the present case the four sets
correspond to $50\, ,133,\, 133,$ and 270 irreducible $\mathbf{k}$-points
for the electronic self-consistency and $793,\, 793,\, 2413,$ and
$5425$ irreducible $\mathbf{k}$-points for the Fermi surface sampling,
respectively. }
\end{figure}

Finally, in Fig. 4 we show the convergence of $\omega _{\mathbf{q}\nu }$,
$\gamma _{\mathbf{q}\nu }$, and $\lambda _{\mathbf{q}\nu }$ at $A$
in the Brillouin zone for $NbB_{2}$ as a function of $\mathbf{k}$-points
using the double-grid technique as outlined in Ref. \cite{savrasov2}.
The use of double-grid technique allows one to construct two separate
but commensurate $\mathbf{k}$-grids, one for the electronic charge
self-consistency and the other for Fermi-surface sampling. We employed
4 sets of double grids (i) (8, 8, 8, 24), (ii) (12, 12, 12 ,24), (iii)
(12, 12, 12, 36), and (iv) (16, 16, 16, 48), where the first three
numbers define the electronic self-consistency grid and the last number
sets up the Fermi-surface sampling grid which is commensurate with
the first grid. We find that the results are converged for the (12,12,12,36)
grid used in the present work.

In conclusion, we have studied from first principles (i) the phonon
dispersion, (ii) the phonon linewidth, and (iii) the partial electron-phonon
coupling constant along $\Gamma -A$ in $MgB_{2},\, NbB_{2}$ and
$TaB_{2}$ in $P6/mmm$ crystal structure. We find that (i) in contrast
to a strong and $E_{2g}$-mode dominated electron-phonon coupling
in $MgB_{2}$, the transition-metal diborides $NbB_{2}$ and $TaB_{2}$
have a relatively weak electron-phonon coupling which is dominated
by the $LA$ mode, (ii) the $E_{2g}$ phonon linewidth is an order
of magnitude larger in $MgB_{2}$ than in $NbB_{2}$ or $TaB_{2}$,
and (iii) the $E_{2g}$ phonon frequency in $NbB_{2}$ as well as
$TaB_{2}$ is considerably higher than in $MgB_{2}$ while the $LA$
phonon frequency at $A$ for $TaB_{2}$ is almost half of that of
$MgB_{2}$ or $NbB_{2}$.

\end{document}